\documentclass[aps,english,prb,amsmath,amsfonts,amssymb,superscriptaddress,twocolumn,showpacs,citeautoscript]{revtex4-1}
\usepackage[T1]{fontenc}
\usepackage[latin9]{inputenc}
\usepackage{amsmath}
\usepackage{amssymb}
\usepackage{wasysym}
\usepackage{esint}
\usepackage{graphicx}
\usepackage{times}
\usepackage{nicefrac}
\usepackage{appendix}
\usepackage{textcase}
\usepackage{subfigure}
\usepackage{empheq}
\usepackage{color}
\usepackage{leftidx}
\usepackage{makecell}
\usepackage{stackrel}
\makeatletter
\@ifundefined{textcolor}{}
{%
 \definecolor{BLACK}{gray}{0}
 \definecolor{WHITE}{gray}{1}
 \definecolor{RED}{rgb}{1,0,0}
 \definecolor{GREEN}{rgb}{0,1,0}
 \definecolor{BLUE}{rgb}{0,0,1}
 \definecolor{CYAN}{cmyk}{1,0,0,0}
 \definecolor{MAGENTA}{cmyk}{0,1,0,0}
 \definecolor{YELLOW}{cmyk}{0,0,1,0}
}

\renewcommand{\vec}[1]{\mathbf{#1}}
\renewcommand{\Re}{\operatorname{Re}}
\renewcommand{\Im}{\operatorname{Im}}

\newcommand{\sym}{\operatorname{sym}}

\renewcommand{\b}{\beta}

\newcommand{\PT}{\mathcal{PT}}

\newcommand{\add}[1]{\if\a\b{{\color{red} #1}}\else{#1}\fi}

\newcommand{\citeasnoun}[1]{Ref.~\onlinecite{#1}}

\renewcommand{\eqref}[1]{(\ref{eq:#1})}

\newcommand{\figref}[1]{Fig.~\ref{fig:#1}}
\newcommand{\Figref}[1]{Figure~\ref{fig:#1}}

\newcommand{\secref}[1]{Sec.~\ref{sec:#1}}

\def\ro{\vec{x}}
\def\rs{\vec{y}}
\newcommand{\mat}[1]{{#1}}

\newcommand{\Tr}{\text{Tr }}

\makeatother

\usepackage{babel}
\begin{document}
\title{Amplified and directional spontaneous emission from arbitrary
  composite bodies: \\ a self-consistent treatment of Purcell effect
  below threshold}

\author{Weiliang Jin}
\affiliation{Department of Electrical Engineering, Princeton University, Princeton, NJ 08544, USA}
\author{Chinmay Khandekar}
\affiliation{Department of Electrical Engineering, Princeton University, Princeton, NJ 08544, USA}
\author{Adi Pick}
\affiliation{Department of Physics, Harvard University, Cambridge, MA 02138, USA}
\author{Athanasios G. Polimeridis}
\affiliation{Skolkovo Institute of Science and Technology, Moscow, Russia}
\author{Alejandro W. Rodriguez}
\affiliation{Department of Electrical Engineering, Princeton University, Princeton, NJ 08544, USA}

\begin{abstract} 
  We study amplified spontaneous emission (ASE) from wavelength-scale
  composite bodies---complicated arrangements of active and passive
  media---demonstrating highly directional and tunable radiation
  patterns, depending strongly on pump conditions, materials, and
  object shapes. For instance, we show that under large enough gain,
  $\PT$ symmetric dielectric spheres radiate mostly along either
  active or passive regions, depending on the gain distribution. Our
  predictions are based on a recently proposed fluctuating
  volume--current (FVC) formulation of electromagnetic radiation that
  can handle inhomogeneities in the dielectric and fluctuation
  statistics of active media, e.g. arising from the presence of
  non-uniform pump or material properties, which we exploit to
  demonstrate an approach to modelling ASE in regimes where Purcell
  effect (PE) has a significant impact on the gain, leading to spatial
  dispersion and/or changes in power requirements. The nonlinear
  feedback of PE on the active medium, captured by the Maxwell--Bloch
  equations but often ignored in linear formulations of ASE, is
  introduced into our linear framework by a self-consistent
  renormalization of the (dressed) gain parameters, requiring the
  solution of a large system of nonlinear equations involving many
  \emph{linear} scattering calculations.
\end{abstract}
\maketitle





Noise in structures comprising passive and active materials can lead
to important radiative effects~\cite{premaratne2011light}, e.g.
spontaneous emission (SE)~\cite{Amplifiedbook},
superluminescence~\cite{dicke1954coherence}, and
fluorescence~\cite{lahoz2013high}. Although large-etalon gain
amplifiers and related devices have been studied for
decades~\cite{allen1973amplified,giles1991modeling,gutierrez2006uni,wang2007high},
there is increased interest in the design of wavelength-scale
composites for tunable sources of scattering and incoherent
emission~\cite{yoo2011quantum,ge2014parity}, or which serve as perfect
absorbers~\cite{chong2010coherent} at mid-infrared and visible
wavelengths.


In this paper, we extend a recently developed fluctuating--volume
current (FVC) formulation of electromagnetic (EM)
fluctuations~\cite{polimeridis2014stable,polimeridis2015computation,polimeridis2015fluctuating,jin2015temperature}
to the problem of modeling spontaneous emission and scattering from
composite, wavelength-scale structures, e.g. metal--dielectric
spasers~\cite{bergman2003surface,stockman2011spaser,stockman2013spaser},
subject to inhomogeneities in both material and noise properties. We
begin by studying amplified spontaneous emission (ASE) from
piecewise-constant composite bodies, showing that their emissivity can
exhibit a high degree of directionality, depending sensitively on the
gain profile and shape of the objects. For instance, we find that
under large enough gain, the directivity of parity-time ($\PT$)
symmetric spheres can be designed to lie primarily along active or
passive regions, depending on the presence or absence of
centrosymmetry, respectively. Such composite micron-scale emitters act
as tunable sources of incoherent radiation, forming a special class of
infrared/visible antennas exhibiting polarization- and
direction-sensitive absorption and emission properties. An important
ingredient for the design of directional emission is the ability to
tune the gain profile of the objects, which can be far from
homogeneous in realistic settings. Here, we consider two important
sources of inhomogeneities affecting population inversion of
atomically doped media: inhomogeneous pump profiles and modifications
stemming from changes to the emitters' local radiative
environment. Below threshold, the latter stems primarily from changes
to atomic decay rates, which can be either enhanced or suppressed
through the Purcell effect (PE)~\cite{iwase2010analysis}. Because PE
is sensitive to the gain and geometry of the objects, such a
dependence manifests as a nonlinear and nonlocal interaction (or
feedback) between the atomic medium and the optical
environment~\cite{gu2013purcell}, which we model within the
stationary--phase approximation~\cite{cerjan2015steady} via a
self-consistent renormalization of the (dressed) atomic parameters. We
show that this leads to a system of nonlinear equations, involving as
many degrees of freedom as there are volumetric unknowns, which in
principle require many scattering problems (radiation from dipoles) to
be solved simultaneously, but which thanks to the low-rank nature of
volume--integral equation (VIE) scattering
operators~\cite{polimeridis2015fluctuating} can be accurately obtained
with far fewer scattering calculations than there are unknowns. Our
predictions indicate that under significant PE, composite objects can
exhibit a high degree of dielectric gain enhancement/suppression and
inhomogeneity, affecting power requirements and emission patterns.

Gain--composite structures are the subject of recent theoretical and
experimental work~\cite{premaratne2011light}, and have been studied in
a variety of different contexts, including spasers (combinations of
metallic and gain media) with low-threshold
characteristics~\cite{bergman2003surface,stockman2011spaser,stockman2013spaser,parfenyev2012intensity,arnold2013dye},
random structures with special absorption
properties~\cite{zaitsev2010recent,andreasen2010numerical,bachelard2012taming,liew2014active,ge2014enhancement},
and nano-scale particles with highly tunable emission and scattering
properties~\cite{fan2010huge,prosentsov2013light}. $\PT$--symmetric
structures have received special attention recently as they shed
insights into important non-Hermitian physics, such as design criteria
for realizing exceptional points~\cite{heiss2012physics},
symmetry--breaking~\cite{heiss2012physics}, uni-directional
scattering~\cite{lin2011unidirectional,ge2012conservation,alaeian2014parity},
and lasing thresholds~\cite{kulishov2013distributed}. Until
recently, 
most studies of radiation/scattering from $\PT$ structures remained
confined to 1d and 2d
geometries~\cite{chong2011p,longhi2014pt,ge2012conservation,alaeian2014parity,miri2014scattering,graefe2011pt,turduev2014asymmetric,yoo2011quantum}. In
such low-dimensional systems, it is common to employ scattering matrix
formulations~\cite{longhi2014pt,ge2012conservation,chong2011p} to
solve for the complex eigenmodes and scattering properties of bodies,
leading to many analytical insights. For instance, while the
introduction of gain violates energy conservation, a generalized
optical theorem can be obtained in 1d, establishing conditions for
unidirectional transmission of
light~\cite{lin2011unidirectional,ge2012conservation,alaeian2014parity}. Other
studies focus on 2d high--symmetry objects such as cylindrical or
spherical bodies~\cite{miri2014scattering,graefe2011pt} or particle
lattices~\cite{turduev2014asymmetric}, demonstrating strong asymmetric
and gain-dependent scattering cross-sections, while 3d structures such
as ring resonators have been studied within the framework of
coupled-mode theory~\cite{feng2014single,peng2014parity}. With few
exceptions~\cite{yoo2011quantum}, however, most studies of
gain--composite bodies have focused on their scattering rather than
emission properties.





Furthermore, while these systems are typically studied under the
assumption of
piecewise-constant~\cite{yoo2011quantum,chong2011p,ge2012conservation}
or linearly varying~\cite{ge2014parity} gain profiles, in realistic
situations, inhomogeneities in the pump or material parameters
(e.g. arising from PE~\cite{gu2013purcell}, hole
burning~\cite{haken1985light,siegman1986lasers}, and gain
saturation~\cite{siegman1986lasers}) result in spatially varying
dielectric profiles which alter SE. For example, the highly-localized
nature of plasmonic resonances in spasers result in strongly
inhomogeneous pumping rates~\cite{stockman2011spaser} and
orders-of-magnitude enhancements in atomic radiative decay
rates~\cite{zhou2013lasing}. In random
lasers~\cite{zaitsev2010recent}, partial pumping plays an important
role in determining the lasing
threshold~\cite{andreasen2011spectral,bachelard2012taming} and
directionality~\cite{liew2014active,ge2014enhancement}. Rigorous
descriptions of lasing effects in these systems commonly resort to
solution of the full Maxwell--Bloch (MB) equations, in which the
electric field $\vec{E}$ and induced (atomic) polarization field
$\vec{P}$ couple to affect the atomic population decay
rates~\cite{cerjan2012steady}. However, the MB equations are a set of
coupled, time-dependent, nonlinear partial differential
equations~\cite{esterhazy2014scalable} which, not surprisingly, prove
challenging to solve except in simple situations involving
high--symmetry~\cite{zhou2013lasing,hess2012active} or low-dimensional
structures~\cite{premaratne2011light}. Recently, brute-force FDTD
methods have been employed to study the transition from ASE to lasing
in 1d random media~\cite{andreasen2010numerical,buschlinger2015light},
2d metamaterials~\cite{pusch2012coherent,hess2012active}, photonic
crystals~\cite{bermel2006active}, and more recently,
nano-spasers~\cite{zhou2013lasing}.

A more recent, general-purpose method that is applicable to arbitrary
structures is the steady-state ab-initio laser theory (SALT), an
eigenmode formulation that exploits the stationary-inversion
approximation to remove the time dependence and internal atomic
dynamics of the MB
equations~\cite{cerjan2015quantitative,cerjan2015steady,cerjan2015laser},
yet captures important nonlinear effects such as hole burning and gain
saturation~\cite{cerjan2015steady} through effective two-level
polarization and population equations~\cite{de2006semiclassical}. The
resulting nonlinear eigenvalue equation can be solved via a
combination of
Newton-Raphson~\cite{esterhazy2014scalable,burkhardt2015steady},
sparse-matrix solver~\cite{henon2002pastix}, and nonlinear
eigenproblem~\cite{guillaume1999nonlinear} techniques, in combination
with either spectral ``CF'' basis expansions (especially suited for
structures with special symmetries)~\cite{tureci2006self} or
brute-force methods~\cite{andreasen2009finite} that can handle a wider
range of shapes and conditions. Although this formulation can describe
many situations of interest, it nevertheless poses computational
challenges in 3d or when applied to structures supporting a large
number of modes~\cite{esterhazy2014scalable}. Furthermore, the impact
of noise below or near threshold has yet to be addressed, although
recent progress is being made along these
directions~\cite{esterhazy2014scalable,burkhardt2015steady}.

Below and near the lasing threshold, however, stimulated emission is
often negligible, enabling linearized descriptions of the gain
medium~\cite{siegman1986lasers,campione2011complex,cerjan2012steady,chua2011low}. Such
approximations, however, ignore nonlinearities stemming from the
induced radiation rate $\sim \vec{E} \cdot \vec{P}$ present in the MB
equations, which captures feedback on the atomic medium due to
amplification or suppression of noise from changes in the local
density of states (also known as Purcell
effect)~\cite{zhou2013lasing,yasumoto2005electromagnetic,lau2008effect}. Here,
we show that PE can be introduced into the linearized framework via a
self-consistent renormalization or dressing of the gain parameters, an
approach that was recently suggested~\cite{gu2013purcell} but which
has yet to be demonstrated. In particular, working within the scope of
the linearized MB equations and stationary-inversion approximation, we
capture the nonlinear feedback of ASE on gain, i.e. the steady-state
enhancement/suppresion of gain and atomic decay rates due to PE, via a
series of nonlinear equations involving many coupled, \emph{linear},
classical scattering calculations---local density of states or
far-field emission due to electric dipole currents. Since the gain
profile can become highly spatially inhomogeneous, it is advantageous
to tackle this problem using brute-force methods, e.g. finite
differences~\cite{taflove1995computational}, finite
elements~\cite{chew2001fast}, or via the scattering VIE framework
described below. Our FVC method is particularly advantageous in that
it is general and especially suited for handling scattering problems
with large numbers of degrees of freedom (defined only within the
volumes of the objects), in contrast to eigenmode
expansions~\cite{tureci2006self} which become inefficient in
situations involving many
resonances~\cite{esterhazy2014scalable,tureci2006self} or near-field
effects~\cite{liu2015near,chinamy2015near}.


\section{Theory}
\label{sec:theory}

\emph{Active medium.---} To begin with, we review the linearized
description of a gain medium consisting of optically pumped 4-level
atoms: a dense collection of active emitters (e.g. dye
molecules~\cite{siegman1986lasers} or quantum
dots~\cite{strauf2011single,gregersen2012quantum}) embedded in a
passive (background) dielectric medium $\epsilon_r$. (Note that our
choice of 4-level system here is merely illustrative since the same
approach described below also applies to other active media.) The
effective permittivity of such a medium is well described by a simple
2-level Lorentzian gain
profile~\cite{siegman1986lasers,campione2011complex,chua2011low},
\begin{equation}
  \epsilon(\omega,\vec{x})=\epsilon_r(\omega)+\underbrace{\frac{4\pi g(\vec{x})^2}{\hbar\gamma_{\perp}}\frac{\gamma_{\perp}D_0(\vec{x})}{\omega-\omega_{21}+i\gamma_\perp}}_{\epsilon_g(\vec{x},\omega)},
  \label{eq:eg}
\end{equation}
where $\epsilon_g$ depends explicitly on the frequency $\omega_{12}$,
polarization decay rate $\gamma_\perp$ (or gain bandwidth), coupling
strength $g^2=\frac{3\hbar
  c^3}{2\sqrt{\epsilon_r}\omega_{21}^3}\gamma_{21}^r$~\cite{chua2011low,campione2011complex},
and inversion factor $D_0 = n_2-n_1$ associated with the $2\to 1$
transition. Under the adiabatic or stationary-inversion
approximation~\cite{cerjan2015steady} and assuming that the system is
pumped at $\omega_{30}$, the steady-state population inversion is
given by:
\begin{align}
  D_0&=\frac{\left(1-\frac{\gamma_{21}}{\gamma_{10}}\right)\mathcal{P}/\gamma_{21}}{1+\left(A\frac{\gamma_{21}}{\gamma_{32}}+\frac{\gamma_{21}}{\gamma_{10}}+1\right)\mathcal{P}/\gamma_{21}}n.
\end{align}
(Note that in a dense medium, $\gamma_\perp \gg \gamma_{21}$ is
dominated by collisional and dephasing
effects~\cite{siegman1986lasers}.)  Here, $A=1 (2)$ for incoherent
(coherent) pump source; $c$ denotes the speed of light, $n_i$ the
population (per unit volume) of level $i$, $n = \sum_i n_i$ the
overall atomic population,
$\gamma_{ij}=\gamma_{ij}^r+\gamma_{ij}^{nr}$ the decay rate from level
$i \to j$, consisting of radiative and non-radiative terms,
respectively, and $\mathcal{P}(\vec{x}) = \frac{\sigma
  c}{\sqrt{\epsilon_r(\omega_{30})}\hbar \omega_{30}}
|E(\ro,\omega_{30})|^2$ is the position-dependent pump rate from $0
\to 3$, which is often the main source of spatial dispersion.

The SE properties of such a medium are described by the
fluctuation--dissipation theorem
(FDT)~\cite{matloob1997electromagnetic,jeffers1993quantum,graham1968quantum}. In
particular, from local thermodynamic considerations, one can show that
associated with the presence of absorption or amplification are
fluctuating polarization currents $\sigma$ whose correlation
functions, $\langle \sigma_i(\ro,\omega) \sigma_j^\ast(\rs,\omega)
\rangle = \frac{4}{\pi} \Im \epsilon_g(\ro,\omega) n_2/(n_1-n_2)
\delta(\ro-\rs)\delta_{ij}$, depend on the corresponding macroscopic
gain profile $\Im \epsilon_g$ and population inversions
$n_2/(n_1-n_2)$. The latter is often described in terms of an
effective (local) temperature $T$ defined with respect to the Planck
spectrum~\cite{matloob1997electromagnetic},
$(e^{\hbar\omega/k_\mathrm{B} T} - 1)^{-1}$, whose value turns
negative~\cite{patra1999excess} for systems with population inversion
$n_2 > n_1$, where $T \to 0^{-}$ in the limit of complete
inversion~\cite{jeffers1993quantum}. The presence of inhomogeneities
in the dielectric function and fluctuation statistics can be a hurdle
for calculations of SE that rely on scattering-matrix
formulations~\cite{bimonte09,Messina11}, but here we exploit a
recently developed FVC formulation based on the VIE method which
captures all of the relevant physics.

\emph{FVC formulation.---} In order to obtain the individual and/or
cummulative radiation from all dipoles within a given object, we
exploit the FVC formulation introduced
in~\citeasnoun{polimeridis2015fluctuating} and summarized here. The
starting point of FVC is the VIE formulation of EM
scattering~\cite{chew2001fast}, describing scattering of an incident,
6-component electric ($\vec{E}$) and magnetic ($\vec{M}$) field
$\phi_{\rm inc} = (\vec{E}; \vec{H})$ from a body described by a
spatially varying $6\times 6$ susceptibility tensor $\chi(\ro)$.
Given a 6-component electric ($\vec{J}$) and magnetic ($\vec{M}$)
dipole source $\sigma = (\vec{J};\vec{M})$, the incident field is
obtained via a convolution ($\star$) with the $6\times 6$ homogeneous
Green's function (GF) of the ambient medium $\Gamma(\ro,\rs)$, such
that $\phi_{\mathrm{inc}}=\Gamma\star\sigma=\int d^3 \rs
\Gamma(\ro,\rs)\sigma(\rs)$. Exploiting the volume equivalence
principle~\cite{chew2001fast}, the unknown scattered fields $\phi_{\rm
  sca} = \Gamma \star \xi$, can also be expressed via convolutions
with $\Gamma$, except that here $\xi=-i\omega\chi\phi$ are the
(unknown) bound currents in the body, related to the total field
inside the body $\phi=\phi_\mathrm{inc} + \phi_\mathrm{sca}$ through
$\chi$. Writing Maxwell's equations in terms of the bound currents, we
arrive at the so-called JM--VIE equation~\cite{polimeridis2014stable}:
\begin{equation}
  \underbrace{\left[\Gamma\star + (i \omega \chi)^{-1}\right]}_{Z} \xi =- (\Gamma
  \star \sigma),
  \label{eq:VIEop}
\end{equation}
whose solution can be obtained by a Galerkin discretization of the
currents $\sigma(\ro)=\sum_n s_n b_n(\vec{x})$ and $\xi(\ro)=\sum_n
x_n b_n(\vec{x})$ in a convenient, orthonormal basis $\{b_n\}$ of $N$
6-component vectors, with vector coefficients $s$ and $x$,
respectively. The resulting matrix expression has the form $x+s=Ws$,
where the VIE matrix $(\mat{W}^{-1})_{m,n}= \langle b_m,b_n + i \omega
\chi (\Gamma \star b_n) \rangle$ and $\langle , \rangle$ denotes the
standard conjugated inner product. Direct application of Poynting's
theorem yields the following expression for the far-field radiation
flux from $\sigma$ (here, a single dipole source embedded within the
volume)~\cite{Jackson98}:
\begin{align}
  \Phi_\sigma= -\frac{1}{2} \Tr \left[D W^* \sym G W\right]
\label{eq:Phi}
\end{align}
where $D=s^*s$ and $G$ are $N\times N$ matrices, with
$\mat{G}_{mn}=\langle b_m,\Gamma\star b_n~\rangle$.

If $\{b_n\}$ is chosen to be a localized basis of unit-amplitude,
i.e. $\langle b_m, b_n\rangle = \delta_{nm}$, volume elements, then
the flux contribution from a given dipole source in the volume
(including different polarizations) $b_n$ is precisely the diagonal
element $-\frac{1}{2}(W^* \sym G W)_{n,n}$. In contrast, the overall
ASE is given by an ensemble-average over all such fluctuating dipole
sources $\Phi = \langle \Phi_\sigma \rangle$, in which case the
elements of the matrix $\mat{D}$, which encodes information about the
current amplitudes, are given by the current--current correlations
$\langle \mat{D}\rangle_{mn}=\int \int d^3\ro d^3\rs \,
b_{m}^\ast(\ro) \langle \sigma(\ro) \sigma^\ast(\rs) \rangle
b_{n}(\rs)$, determined by the FDT above. Direct computation of
\eqref{Phi} is expensive due to the large dimensionality $N$ of the
problem, but it turns out that the Hermitian, negative-semidefinite,
and low-rank nature of $\sym G$ (since it is associated with the
smooth, imaginary part of the Green's functions) enables re-expressing
the trace as the Frobenius norm of a low-rank matrix. Specifically,
decomposing $\sym G=-U_rS_rU_r^{*}$ via a fast approximate
SVD~\cite{hochman2014reduced}, where $r\ll
N$~\cite{chai2013theoretical}, and further decomposing
$S_r=L_SL_S^{*}$, we find that the product $W^* \sym G W$ can be
written in the form $Q Q^{*}$, with $Q=W^{*}U_rL_S$, reducing the
calculation of the diagonal elements to a small series $r \ll N$ of
scattering calculations (matrix-inverse
operations)~\cite{polimeridis2014stable}. Similarly, as shown in
\citeasnoun{polimeridis2015fluctuating}, the same follows for the
overall ASE which is just a weighted sum of the diagonal elements. For
example, while the calculations below require $N\geq 40^3$ basis
functions to obtain accurate spatial resolution, we find that
generally $r \lesssim 20$.

\begin{figure*}[t!]
\begin{center}
\includegraphics[width=2\columnwidth]{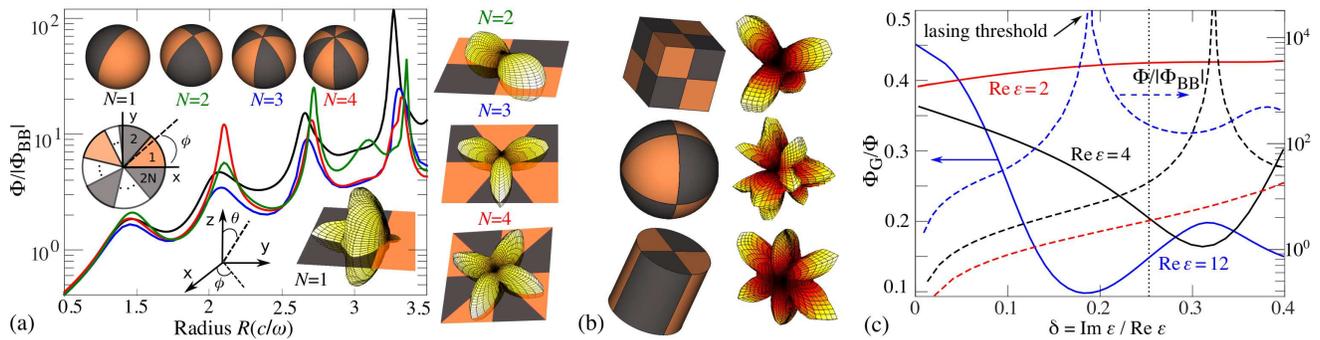}
\caption{(a) SE flux $\Phi$ from $\PT$--symmetric spheres consisting
  of $N=\{1,2,3,4\}$ (black, green, blue, and red lines) regions of
  equal gain (red) and loss (blue), with permittivities $\epsilon =
  4\pm i$ and gain temperature $T \to 0^{-}$~K (corresponding to
  complete population inversion), as a function of sphere radius $R$
  (for a fixed vacuum wavelength $2\pi c/\omega$). $\Phi(\omega)$ is
  normalized by the flux $|\Phi_\mathrm{BB}(\omega)| = R^2
  (\omega/c)^2\hbar \omega$ from a ``blackbody'' of the same surface
  area and temperature. Insets show angular radiation intensities
  $\Phi(\theta,\phi)$ for different $N$ at selective radii
  $R=\{2.6,3,2.7,2.1\} (c/\omega)$ (corresponding to increasing $N$),
  with orange/gray colormaps denoting regions of gain/loss. (b)
  Selected $\Phi(\theta,\phi)$ of various $\PT$--symmetric shapes,
  including ``beach balls'', ``magic'' cubes, and cylinders, showing
  complex directivity patterns that depend strongly on the
  centro-symmetry of the objects. Here the color white (black) refers
  to the maximum (minimum) flux value. (c) Peak SE flux $\Phi$ (right
  axis, dashed lines) and directivity $\Phi_G/\Phi$ (left axis, solid
  lines), or the ratio of the flux emitted along gain direction
  $\Phi_G$ (see text) to the total flux $\Phi$, near the third
  resonance of the $N=1$ sphere, as a function of the gain/loss
  tangent $\delta = |\Im \epsilon|/\Re \epsilon$ and for multiple
  values of $\Re \epsilon = \{2,4,12\}$ (red, black, and blue lines),
  where the vertical dashed black line denotes $\delta$ for (a). The
  plots show increased directivity attained as the system approaches
  the lasing threshold (marked in the case of $\Re\epsilon=12$, where
  $\Phi$ diverges ).}
\label{fig:uniform}
\end{center}
\end{figure*}

\emph{Purcell effect.---} Although often assumed to be uniform, the
atomic radiative decay rates $\gamma_{ij}^r$ entering \eqref{eg} are
in fact position dependent due to
PE~\cite{yasumoto2005electromagnetic,zhou2013lasing}, leading to
changes in the dipole coupling $g$ and population inversion factor
$D_0$, either enhancing or suppressing (quenching)
gain~\cite{yasumoto2005electromagnetic}.  In what follows, we only
consider modifications to the radiative decay rate at the lasing
transition $\gamma_{21}^r$; unlike the pump which is incident at a
non-resonant frequency, changes in $\gamma_{21}^r$ can have a
significant impact on SE and must therefore be treated
self-consistently.

The impact of PE on the radiative decay rate $\gamma^r_{21}$ of an
atom at some position $b_n$ is captured by the coupling of the atomic
polarization and electric fields $\sim \vec{P}\cdot\vec{E}$, or the
induced radiation term in the MB equations, in the presence of the
noise and surrounding dielectric
environment~\cite{waks2010cavity}. While technically this requires
abandoning the linear model above, the weak nature of noise (ignoring
stimulated emission) implies that the latter can also be obtained
(perturbatively) from a \emph{linear}, classical calculation: the
radiative flux $\Phi_{b_n}$ from a classical dipole at
$b_n$. Specifically, the renormalized or dressed decay rate of an atom
at position $\vec{x}$ can be expressed
as~\cite{yasumoto2005electromagnetic}:
\begin{equation} 
  \gamma^r_{21}(\vec{x}) = \mathcal{F}(\vec{x}) \gamma^{r,0}_{21},
\label{eq:gamma}
\end{equation}
where $\mathcal{F}(\vec{x})$ denotes the Purcell factor of a dipole at
$\vec{x}$, and the supper-script ``0'' denotes the decay rate of the
atomic population in the lossy (background) medium. It follows that
the decay rate associated with a given $b_n$ and entering \eqref{eg}
is given by $\gamma_{21}(b_n)=\gamma_{21}^{nr}+\mathcal{F}_{b_n}
\gamma_{21}^{r,0}$, where the Purcell factor,
\begin{equation}
  \mathcal{F}_{b_n}=\Phi_{b_n}/\Phi^0.
\end{equation}
(Note that all $\Phi_{b_n} \sim (W^* \sym G W)_{n,n}$ can be computed
very fast, as explained above.) Here, we assume that the bulk
(background) medium $\epsilon_r$ only has a significant impact on
$\gamma^{nr,0}_{21}$~(obtained either experimentally or theoretically
by accounting for atomic interactions within the
bulk)~\cite{siegman1986lasers} but not on the radiative decay rate
$\gamma^{r,0}_{21}$, in which case $\Phi^0 = \omega^4/12\pi\epsilon_0
c^3$ is the emission rate of the atom in vacuum (assuming a
unit-amplitude dipole). Note that in a lossy medium, e.g. in metals,
the bare $\gamma^{0}_{21}$ will be dominated by non-radiative
processes~\cite{wu2009high,fang2012plasmon}, leading to small quantum
yields (QY) $\gamma^{r,0}_{21}/\gamma^{0}_{21} \ll 1$.  Note also that
technically, the calculation of the Purcell factor requires
integration over the gain bandwidth, $\mathcal{F}_{b_n} = \int d\omega
\frac{\gamma_\perp/\pi}{\gamma_\perp^2 + (\omega-\omega_{21})^2}
\frac{\Phi_{b_n}(\omega)}{\Phi^0(\omega)}$, but here we make the
often-employed and simplifying assumption that $\gamma_\perp \gg
\gamma_{21}$ and $\gamma_\perp \lesssim$ spectral radiative
features~\cite{lu2014enhancing,kristensen2013shell}, so as to only
consider radiation at $\omega_{21}$.

The gain profile of a body subject to an incident pump rate
$\mathcal{P}$ can be obtained by enforcing that \eqref{eg} and
\eqref{gamma} be satisfied simultaneously. Such systems of nonlinear
equations are most often solved iteratively using one or a combination
of algorithms, ranging from simple fixed-point
iteration~\cite{suli2003introduction} to more sophisticated approaches
like Newton--Raphson and nonlinear Arnoldi
methods~\cite{esterhazy2014scalable,voss2004arnoldi}. Essentially,
starting with the bare parameters, dressed decay rates are computed
via \eqref{gamma} from the radiation equation \eqref{Phi} after which,
having updated the gain--medium equation \eqref{eg}, the entire
process is repeated until one arrives at a fixed-point of the
system. In principle, this requires hundreds of thousands $N$ of
scattering calculations (flux from each dipole source in the active
region) to be solved \emph{per iteration}, which becomes prohibitive
in large systems, but the key here is that the entire spatially
varying flux $\{\Phi_{b_n}\}$ throughout the body can be computed
extremely fast, requiring far fewer ($\ll N$) scattering calculations
(as described above).  Note that these large systems of nonlinear
equations have many fixed points and hence convergence to the correct
solution is never guaranteed, depending largely on the inital guess
and algorithm employed~\cite{walker2011anderson}. However, a
convenient and effective approach is to begin by first solving the
system in the fast-converging (passive) regime $\mathcal{P} n/
\gamma_{12} \ll 1$, and then employing this solution as an initial
guess at larger pumps.

\section{Results}


We begin this section by showing that wavelength-scale composite
bodies can exhibit highly complex, tunable, and directional radiation
patterns. Although it is not surprising that objects undergoing ASE
(once known as ``mirrorless lasers'') exhibit highly directed
radiation patterns~\cite{ramachandran2002mirrorless}, few studies have
gone beyond large-etalon Fabry-Perot cavities~\cite{zhu2013amplified}
or fiber waveguides~\cite{harrington2004infrared}, often modelled via
ray-optical or scalar-wave equations~\cite{premaratne2011light}, which
miss important effects present in wavelength-scale
systems~\cite{esterhazy2014scalable}. Further below, we show that
dielectric inhomogeneities arising from the pump and/or radiation
process can also introduce important changes to the ASE patterns. In
particular, we apply the renormalization approach described above to
consider the \emph{nonlinear} impact of Purcell effect on the gain
medium, and show that while in many cases a homogeneous approximation
leads to accurate results, there are situations where these can fail
dramatically. Our calculations are only meant to serve as proof of
principle and revolve around highly doped (Er$^{3+}$ and Rhodamine)
but simple dielectric objects, allowing faster computations but
requiring very large values of $\Im \epsilon_g$ to achieve significant
gain. Similar results follow, however, in systems subject to smaller
$\epsilon_g$ or smaller doping densities, at the expense larger pump
powers or by exploiting resonances with greater confinement or smaller
radiative loss rates (e.g. compact bodies of larger dimensions and/or
refractive indices, or more complex structures such as
photonic-crystal resonators).

\emph{Tunable radiation patterns.---} We begin by exploring SE from
piecewise-constant $\PT$--symmetric spheres [\figref{uniform} insets]
consisting of a background dielectric medium, e.g. nano-composite
polymers~\cite{zhang2012embedding,macdonald2015intrinsic} or
semiconductors~\cite{eliseev1995semiconductor,feng2014single}, doped
with active materials to realize different ($N$) regions of equal gain
(red) or loss (blue). \Figref{uniform}(a) shows the SE flux $\Phi$
from spheres of varying $N = \{1,2,3,4\}$ (black, green, blue, red
lines) and gain/loss permittivities $\epsilon = 4\pm i$, at a fixed
frequency $\omega$ and gain temperature $T \to 0^{-}$~K (corresponding
to complete inversion), as a function of radius $R$ (in units of the
vacuum wavelength $c/\omega$). $\Phi(\omega)$ is normalized by the
flux $|\Phi_\mathrm{BB}(\omega)| = R^2 (\omega/c)^2\hbar \omega$ from
a ``blackbody'' of the same surface area and temperature.  As
expected, $\Phi$ exhibits peaks at selected $R \gtrsim (c/\omega)$
corresponding to enhanced emission at Mie resonances. (Note that peaks
in $\Phi$ continue to increase in amplitude with increasing $R$, due
to decreased radiative losses, with the bandwidth of the resonances
narrowing as the system reaches the lasing transition, at which point
our linear approach breaks down.) Associated with increased ASE is
increased directivity, illustrated by the radiation patterns
$\Phi(\theta,\phi)$ shown on the insets of \figref{uniform}(a) at
selected $R$, whose high directionality contrast sharply with the
emission profile of passive particles. (With few
exceptions~\cite{jin2015temperature}, the latter tend to emit
quasi-isotropically, as can be verified by decreasing the gain of the
spheres.) We find that the direction of largest ASE changes
drastically with respect to $N$, with radiation coming primarily from
either active or passive regions depending on whether the spheres
exhibit or lack centrosymmetry, respectively (insets). In particular,
the ratio $\Phi_G/\Phi$ of the flux emitted from the gain surfaces,
\[
\Phi_G = \sum_{k=0}^{N-1} \int d\theta \sin^2(\theta) \int_{2\pi
  k/N}^{\pi (2k+1) / N} d\phi \, \Phi(\theta,\phi),
\]
to the total flux $\Phi$, is generally $\lesssim 0.5$ for odd $N$
(centrosymmetric) and $\gtrsim 0.5$ otherwise. For instance, $N=1$
spheres exhibit $\Phi_G/\Phi \approx 0.1$ at $R \approx 3.3
(c/\omega)$ whereas $N=2$ spheres exhibit $\Phi_G/\Phi \approx 0.75$
at $R \approx 3 (c/\omega)$. The sensitive dependence of emission
pattern on geometry and gain profile is not unique to spherical
structures, as illustrated in \figref{uniform}(b), which shows
$\Phi(\theta,\phi)$ for various shapes, including ``magic'' cubes,
``beach balls'', and cylinders---as before, the presence/absence of
centro-symmetry results in high/low gain directivity.

To understand the features and origin of these emission patterns,
\figref{uniform}(c) explores the dependence of the peak $\Phi$ (dashed
lines) and $\Phi_G/\Phi$ (solid lines) on the gain/loss tangent
$\delta = |\Im \epsilon|/\Re \epsilon$ of the $N=1$ sphere, near the
third resonance and for multiple values of $\Re
\epsilon=\{2,4,12\}$. As shown, there is negligible ASE in the limit
$\Im \epsilon \to 0$, yet the localization of fluctuating dipoles to
the gain-half of the (increasingly uniform) sphere leads to a small
(though observable) amount of directionality, favoring emission toward
the loss direction. The tendency of dipoles within a sphere to emit in
a preferred direction has been studied in the context of
fluorescence~\cite{lock2001semiclassical} in the ray-optical limit $R
\gg c/\omega$, which as shown here is exacerbated in the presence of
gain~\cite{chen2012amplified}: essentially, dipoles within a sphere
tend to emit in the direction opposite the nearest surface, which
explains why spheres having/lacking centro-symmetry tend to emit along
directions of gain/loss. Moreover, in order to achieve large
directivity, there needs to be a significant amount of mode
confinement and gain, as illustrated by the negligible ASE and
directivity of the $\Re \epsilon=2$ sphere. Finally, we find that for
large enough $\Re \epsilon$, the directivity increases with increasing
$\Im \epsilon$, peaking at a critical $\delta$, corresponding to the
onset of lasing. Such a transition is marked by a diverging $\Phi$
near the threshold along with a corresponding narrowing of the
resonance linewidth (not shown). (Note that our predictions close to
and above this critical gain are no longer accurate since they neglect
important effects stemming from stimulated
emission~\cite{grynberg2010introduction}. For instance, at and above
the critical gain, the resonance linewidth goes to zero and then
broadens with increasing $\Im \epsilon$, while it is well known that
nonlinear gain saturation results in a finite laser
linewidth~\cite{cerjan2015quantitative,cerjan2015laser}) Nevertheless,
our results demonstrate that a significant amount of directivity can
be obtained below the onset of lasing, where the linear approximation
is still valid.

\begin{figure}[t!]
\begin{center}
\includegraphics[width=0.9\columnwidth]{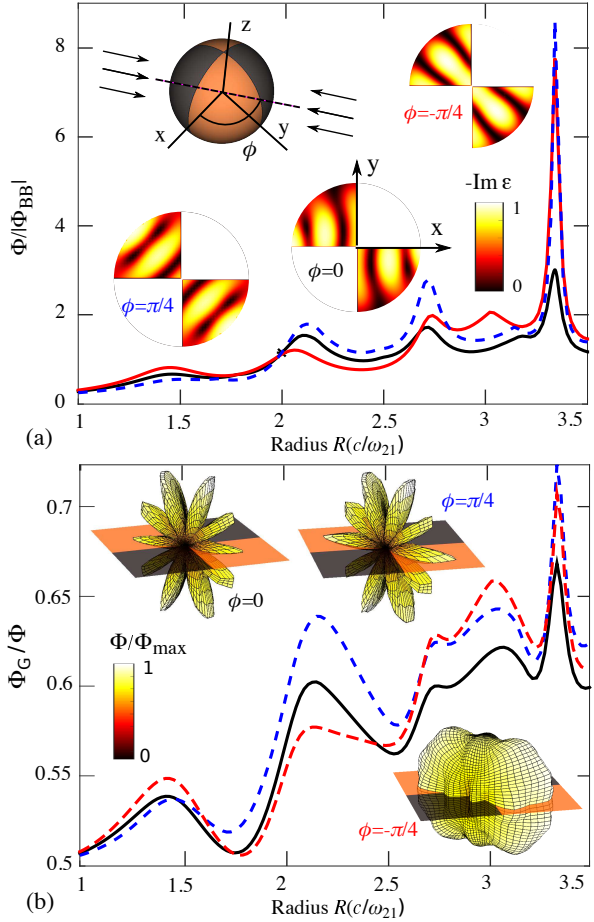}
\caption{(a) SE flux $\Phi$ and (b) gain directivity $\Phi_G/\Phi$, or
  the ratio of the flux emitted along the gain directions $\Phi_G$
  (see text) to the total flux $\Phi$, from the $N=2$ sphere (inset)
  of \figref{uniform} at frequency $\omega_{21}$, corresponding to the
  transition frequency of an active region consisting of Rhodamine 800
  dye molecules. The gain medium is excited by planewaves propagating
  in opposite directions, $\vec{x}\cos\phi+\vec{y}\sin\phi$ and
  $-\vec{x}\cos\phi-\vec{y}\sin\phi$, for three different
  orientations, $\phi=0$ (black line), $\pi/4$ (blue line), and
  $-\pi/4$ (red line), leading to significant spatial
  inhomogeneities. $\Phi$ is normalized by $\Phi_\mathrm{BB}$ as in
  \figref{uniform} and plotted as a function of radius $R$, in units
  of the dye $2\to 1$ transition wavelength $c/\omega_{21}$ (see
  text).  Insets in (a) show $z=0$ cross-sections of the resulting
  gain profiles $-\mathrm{Im} \epsilon_g$ at $R=3.4 (c/\omega_{21})$
  while those in (b) show angular radiation intensities
  $\Phi(\theta,\phi)$ normalized by the maximum intensity $\Phi_{max}$
  at $R=4 (c/\omega_{21})$ for the different pump orientations.}
\label{fig:illumination}
\end{center}
\end{figure}

\begin{figure*}[t!]
\begin{center}
\includegraphics[width=2\columnwidth]{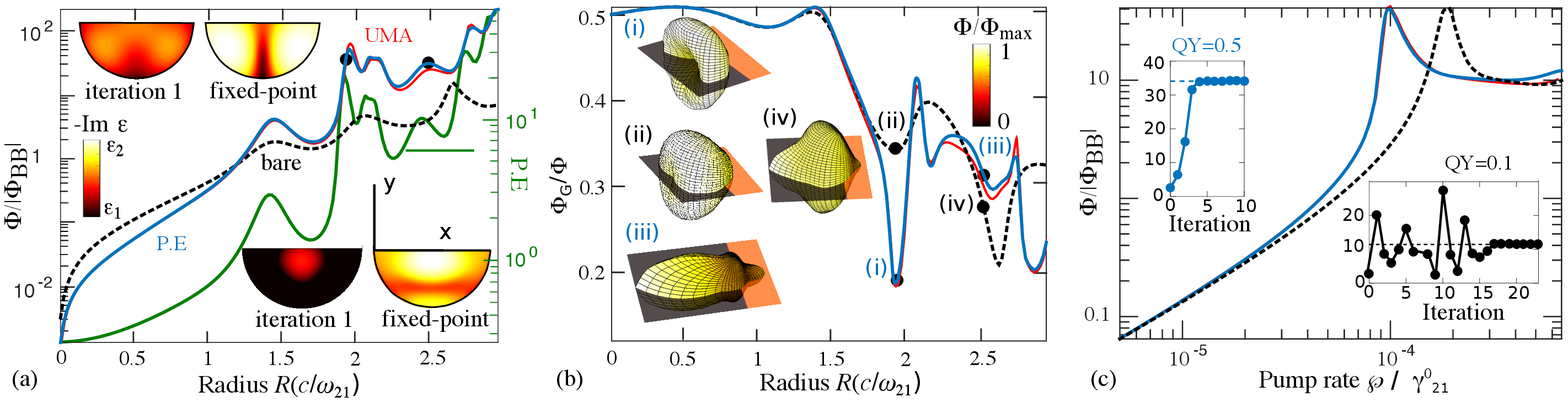}
\caption{Self-consistent treatment of Purcell effect (PE) based on
  solution of \eqref{eg} and \eqref{gamma} using fixed-point
  iteration. (a) SE flux $\Phi$ normalized by $\Phi_\mathrm{BB}$ as in
  \figref{uniform} and (b) gain directivity $\Phi_G/\Phi$ from the
  $N=1$ sphere of \figref{uniform}, as a function of radius $R$ (in
  units of $c/\omega_{21}$), with an active medium consisting of
  Er$^{3+}$ atoms and subject to a uniform pump rate
  $\mathcal{P}/\gamma^0_{21} = 10^{-4}$, where $\omega_{21}$ and
  $\gamma^0_{21}$ denote the bulk (bare) $2\to 1$ transition frequency
  and decay rate of the atoms, respectively.  Top/bottom insets in (a)
  show $z=0$ cross-sections (contour plots) of the gain profiles
  $-\mathrm{Im} \epsilon_g$ obtained during the first and final
  (fixed-point) iteration of the algorithm at two separate radii
  $R=\{2,2.5\} (c/\omega_{21})$, with black/white denoting $\min,\max
  -[\Im \epsilon_g] \approx \{1.9,1.3\}$ and $\{1.9,1.5\}$,
  respectively. Insets in (b) show the angular radiation intensities
  $\Phi(\theta)$ at selected radii (black dots), normalized by the
  maximum intensity $\Phi_{max}$. (c) $\Phi$ at a fixed radius $R = 2
  (c/\omega_{21})$ as a function of $\mathcal{P}/\gamma_{21}^0$, with
  insets illustrating the evolution of $\Phi$ as a function of
  iteration at a fixed $\mathcal{P}/\gamma^0_{21} = 10^{-4}$ and for
  two different values of quantum yields (QY) [see text]. All plots
  compare quantities in the presence (solid blue lines) or absence
  (dashed black lines) of PE, and under a uniform--medium
  approximation (UMA) (solid red lines) described in the text; the
  solid green line in (a) denotes the average Purcell factor
  $\langle\mathcal{F}\rangle = \frac{1}{V}\int_V \mathcal{F}$ at each
  radius.}
\label{fig:PE}
\end{center}
\end{figure*}

\emph{Pump inhomogeneity.---} Next, we employ the 4-level gain model
of \eqref{eg} to illustrate the impact of gain inhomogeneities on
ASE. While both $\mathcal{P}$ and PE are simultaneous sources of
spatial dispersion, for the sake of comparison we consider each
independently of one another.  We begin by studying the impact of pump
on the $N=2$ sphere (above) for an active region consisting of a
background medium $\epsilon_r=4$ that is doped with Rhodamine 800 dye
molecules with atomic parameters: $\omega_{21} = 2.65\times
10^{15}s^{-1}$ ($\lambda\approx 711$~nm),
$\gamma_{\perp}/\omega_{21}=0.04$ , $\gamma_{21}/\omega_{21}=7.5\times
10^{-7}$, $\gamma_{32}/\omega_{21}=\gamma_{10}/\omega_{21}=10^{-3}$,
QY of 20\%, and concentration $n=40$~mM ($2.4\times 10^{19}$
cm$^{-3}$)~\cite{campione2011complex}. Note that since $\omega_{30}
\gg \omega_{21}$, it is safe to neglect feedback due to PE and hence
$\mathcal{P}$ is determined from a single scattering
calculation~\cite{polimeridis2015fluctuating}. For these
parameters~\cite{campione2011complex}, a pump rate
$\mathcal{P}/\gamma_{21} \approx 1$ results in
$-\Im\epsilon_g(\omega_{21})\approx 1$.  We consider illumination with
$\vec{z}$-polarized planewaves incident from two opposite directions
along the $x$--$y$ plane, shown schematically in
\figref{illumination}. The insets depict $z=0$ cross-sections of the
resulting $\epsilon_g$ profile [\figref{illumination}(a)] at
$R=3.4(c/\omega_{21})$ along with emission patterns
$\Phi(\theta,\phi)$ [\figref{illumination}(b)] at
$R=4(c/\omega_{21})$, under three incident conditions, corresponding
to different directions of incidence,
$\hat{\vec{x}}\cos\phi+\hat{\vec{y}}\sin\phi$ and
$-\hat{\vec{x}}\cos\phi-\hat{\vec{y}}\sin\phi$, with $\phi = \{0,\pm
\pi/4\}$; in each case, the incident power is chosen such that
$\mathrm{max}\{-\mathrm{Im}[\epsilon_g]\}=1$. As shown, the gain
profiles vary dramatically with respect to position and incident
angle, with $\Im\epsilon_g$ changing from $0\to 1$ on the scale of the
wavelength. These spatial variations lead to correspondingly large
changes in the overall ASE [\figref{illumination}(a)] and directivity
[\figref{illumination}(b)]. More importantly, we find that these
features cannot be explained by naive, uniform--medium approximations
(UMA). For instance, replacing $\epsilon_g$ with the average gain
$\langle\epsilon_g\rangle = \frac{1}{V}\int_V \epsilon_g$ in the case
$\phi=-\pi/4$, we find that UMA predicts an emission rate
$\Phi/\Phi_\mathrm{BB}\approx 30$ that is three times larger than that
predicted by exact calculations. Differences in illumination angle
also result in different angular radiation patterns
$\Phi(\theta,\phi)$. For instance, we find that $\phi=-\pi/4$ leads to
much more isotropic radiation than $\phi=\{0,\pi/4\}$, a consequence
of the larger $\Im \epsilon_g$ near the center of the sphere and the
fact that dipoles near the center tend to radiate more isotropically
and efficiently than those which are farther laying farther away.

\emph{Purcell effect.---} We now consider inhomogeneities arising from
PE, assuming a uniform pump and doping concentration. In particular,
we apply the self-consistent framework described in \secref{theory} to
study ASE from the $N=1$ sphere above, but with an active region
consisting of a background medium $\epsilon_r=4$ that is doped with
Er$^{3+}$ atoms~\cite{godard2007infrared,cerjan2012steady}, with
parameters: $\omega_{21}=6.28\times 10^{14}s^{-1}$ ($\lambda \approx
2.8\mu$m), $\gamma_{\perp}/\omega_{21}=0.03$,
$\gamma_{21}/\omega_{21}=5\times 10^{-5}$,
$\gamma_{32}/\omega_{21}=\gamma_{10}/\omega_{21}=1$, bare QY of 50\%,
and concentration $n = 10^{19}$ cm$^{-3}$. (Further below we also
consider a different geometry, a metal-dielectric spaser consisting of
similar gain parameters but passive metallic regions.)  Here, a pump
rate $\mathcal{P}/\gamma_{21}=10^{-4}$ results in $-\Im
\epsilon_g(\omega_{21})\approx 1$ in the absence of PE. As discussed
above, we employ fixed-point iteration to solve \eqref{Phi} and
\eqref{gamma} and hence obtain consistent values of $\epsilon_g$ and
$\Phi$, starting with the bare ($\mathcal{F}=1$) atomic parameters and
iterating until the gain parameters converge to the nearest fixed
point. Generally, the convergence rate of the fixed-point algorithm
depends sensitively on the chosen parameter regime, requiring larger
number of iterations with decreasing $\frac{\partial
  \epsilon_g}{\partial \gamma_{21}^r}$ (decreasing local
slope)~\cite{suli2003introduction}. The convergence also depends on
the degree of nonlinearity in the system, which in the case of our
4-level system can be significant under small QY
($\gamma^r_{21}/\gamma_{21} \ll 1$), large $\mathcal{P}$, or
$\gamma_{21}/\gamma_{10} \lesssim 1$ (in which case there is
significant gain saturation). Nevertheless, in practice we find that
for a wide range of parameters, a judicious combination of fixed-point
iteration and Anderson acceleration~\cite{walker2011anderson} ensures
convergence within dozens of iterations. The bottom/top insets of
\figref{PE}(c) demonstrate the iterative process at a fixed $R = 2
(c/\omega_{21})$ and for two different sets of concentrations
$n=\{1,5\}\times 10^{19}$~cm$^{-3}$ and quantum yields~$\approx
\{10,50\}\%$.

\Figref{PE} illustrates the impact of PE on the emission of the
sphere, showing variation in (a) SE flux $\Phi(\omega_{21})$ and (b)
gain directivity $\Phi_G/\Phi$ of the sphere with respect to radius
$R$ at a fixed $\mathcal{P}/\gamma^0_{21}=10^{-4}$, or with respect to
(c) pump rate $\mathcal{P}$ at a fixed $R=2 (c/\omega_{21})$, both
including (solid blue lines) and excluding (dashed black lines) PE. As
before, $\Phi$ is normalized by $\Phi_\mathrm{BB}$. Shown as insets in
\figref{PE}(a) are $z=0$ cross sections of $\Im \epsilon_g$ for the
first and final (fixed-point) iteration of the algorithm, at two
different radii $R = \{2,2.5\} (c/\omega_{21})$ (black dots),
demonstrating large gain enhancement and spatial variations. As
expected, $\Phi$ is either enhanced or suppressed depending on the
average PE (green line) which we have defined as $\langle \mathcal{F}
\rangle \neq 1$, where for convenience we have defined:
\[
\langle \mathcal{F}\rangle = \frac{1}{V}\int_V \mathcal{F} =
\frac{1}{N}\sum_n \Phi_{b_n} / \Phi^0 = -\frac{1}{2N \Phi^0} \Tr W^*
\sym G W.
\]
(As discussed above, $\langle \mathcal{F}\rangle \sim \Phi$ turns out
to be the Frobenius norm of a low-rank matrix and is therefore
susceptible to fast computations.) As shown, at small $R \lesssim 1.2
(c/\omega_{21})$, or in the absence of resonances,
$\langle\mathcal{F}\rangle<1$ and hence $\Phi$ is suppressed with
respect to the predictions of the bare. Conversely,
$\langle\mathcal{F}\rangle>1$ near resonances and hence $\Phi$ is
enhances. Note that for our choice of parameters, the gain profile
scales linearly with the quantum yield, i.e. $\epsilon_g\propto
\mathrm{QY}=\gamma_{21}^r/\gamma_{21}$, such that in the limit as
$\langle \mathcal{F} \rangle \to \infty$ (ignoring quenching occurring
as $\gamma_{21}^r\to\gamma_{10}$), $-\Im \epsilon_g \to 2$. (For
smaller $\mathrm{QY}\ll 1$, $\epsilon_g$ can be many times larger than
the bare permittivity with increasing $\langle \mathcal{F}\rangle$,
saturating at much larger values of PE.)  In addition to changing the
overall SE rate, PE also modifies the sphere's directivity. This is
illustrated in \figref{PE}(b), which shows enhancements in
$\Phi_G/\Phi$ and correspondingly changes in emission patterns
(insets) at selective $R = \{2,2.5\} (c/\omega_{21})$.

\Figref{PE}(c) also explores the dependence of $\Phi$ on $\mathcal{P}$
at a fixed $R = 2 (c/\omega_{21})$, showing that $\Phi$ peaks at a
finite value of $\mathcal{P}/\gamma^0_{12} \gtrsim 10^{-4}$ and then
decreases with increasing $\mathcal{P}$; the same is true for
$\langle\mathcal{F}\rangle$ and $\langle\epsilon_g\rangle$ (not
shown). Such a non-monotonicity stems from the fact that near the
critical pump rate, $\Im \epsilon_g \approx \Re \epsilon_g$, causing
the resonance frequency to shift to smaller radii, a trend that is
observed both in the presence and absence of PE. Surprisingly,
however, we find that while PE causes large inhomogeneities in
$\epsilon_g$, in both scenarios the peak emission is approximately the
same, suggesting the possibility that one could explain the impact of
PE by a simple UMA.  In what follows, we exploit a UMA that not only
greatly simplifies the calculation of PE but also leads to accurate
results over a wider range of parameters. In particular, we consider a
UMA in which the otherwise inhomogeneous gain profile of the object is
replaced with that of a uniform medium $\epsilon_g(\vec{x}) \to
\langle \epsilon_g \rangle$ (assuming a uniform pump rate) given by
\eqref{eg} but with $\gamma^r_{21} \to \langle \gamma^r_{21}\rangle =
\langle \mathcal{F} \rangle \gamma^{r,0}_{21}$, corresponding to a
homogeneous broadening/narrowing of the gain atoms throughout the
sphere. Within this approximation, the system of nonlinear equations
above is described by a single (as opposed to $N \gg 1$) degree of
freedom $\langle \gamma^r_{21}\rangle$, enabling faster convergence
along with application of algorithms that are especially suited for
handling low-dimensional systems of
equations~\cite{conn2009introduction}. Ignoring other sources of
inhomogeneity (e.g. induced by density or pump variations), such an
approximation allows calculation of $\Phi$ via scattering formulations
best-suited for handling piecewise-constant dielectrics, including
SIE~\cite{RodriguezReid12:FSC} and related scattering
matrix~\cite{bimonte09} methods. The solid red lines in \figref{PE}
are obtained by employing the UMA, demonstrating its validity over a
wide range of parameters. Surprisingly, we find that this holds even
in regimes marked by strong gain saturation (e.g. $\gamma_{10}
\lesssim \gamma_{21}$). It follows that in this geometry, the effect
of PE on radiation can be attributed primarily to the presence of a
larger average gain or pump rate in the sphere, whereas the actual
spatial variation in $\epsilon_g$ is largely unimportant.

\begin{figure}[t!]
\begin{center}
\includegraphics[width=0.9\columnwidth]{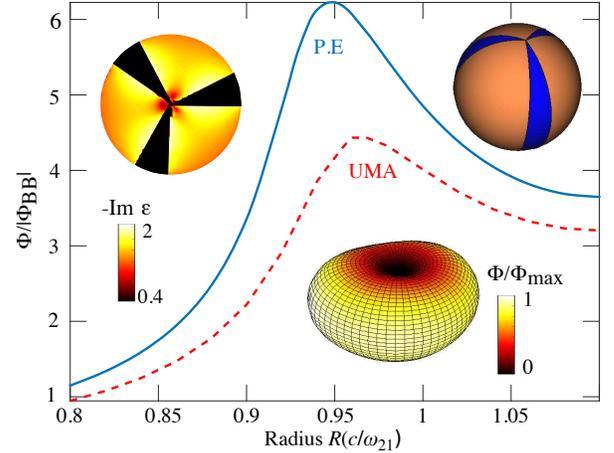}
\caption{Self-consistent treatment of Purcell effect (PE) obtained
  from solution of (1) and (5) (solid blue lines) or by the
  uniform-medium approximation (dashed red line) described in the
  text. SE flux $\Phi$ from a metal--dielectric composite (top-right
  inset) as a function of radius $R$ (in units of $c/\omega_{21}$),
  with orange and blue denoting Er$^{3+}$--doped polymer and Au
  regions, respectively.  Insets show a $z=0$ cross-section of the
  gain profile $-\mathrm{Im} \epsilon_g$ (top) and the angular
  radiation intensity $\Phi(\theta,\phi)$ (bottom) at $R=0.95
  (c/\omega_{21})$ normalized by the maximum intensity $\Phi_{max}$.}
\label{fig:approx}
\end{center}
\end{figure}

There are geometries and situations where such a UMA is expected to
fail, e.g.  structures subject to even large dielectric
inhomogeneities (as in \figref{illumination}). Such conditions arise
in large objects (supporting higher-order resonances) or in
metal--dielectric composites (supporting highly localized surface
waves). \Figref{approx} shows $\Phi$ for one such structure [bottom
inset]: a dielectric sphere with the same gain medium (orange) of
\figref{PE} but partitioned into three metallic (red) regions along
the azithmutal direction, given by $\epsilon(\phi) = -2+i$ for $\phi
\in [2n\pi/3,2n\pi/3+\pi/8]$, where $n=0,1,2$. (Note that our choice
of $\epsilon$ for the metal does not lead to a strong plasmonic
resonance, but still yields significant sub-wavelength confinement.) A
$z=0$ cross-section of the resulting $\epsilon_g$ at
$R=0.95c/\omega_{21}$ is shown on the top inset, illustrating
complicated variations in the gain, whose largest value is attained
near the metal and very rapidly decays within the dielectric.
Comparing the exact (solid blue lines) and UMA (dashed red line)
predictions, one finds that the presence of multiple nodes in
$\epsilon_g$ leads to a dramatic failure for UMA (with a peak error of
$\approx 50\%$). Despite the different radiation rates, however, we
find that UMA effectively captures the main features of the far-field
radiation pattern (inset).

\section{Concluding Remarks}

We have shown that wavelength-scale, active--composite bodies can lead
to complex radiative effects, depending sensitively on the arrangement
of gain and loss. By exploiting a general--purpose formulation of EM
fluctuations, we quantified the non-negligible impact that dielectric
and noise inhomogeneities can have on emission in these
systems. Furthermore, we introduced an approach that captures feedback
from Purcell effect (i.e. the optical environment) on the gain
medium. We note that in situations where ASE is dominated by
relatively few leaky resonances, it is possible and practical to
perform a similar procedure by expanding the fields in terms of
eigenmodes, in which case the problem boils down to solution of a
linear generalized eigenvalue problem for the leaky modes. (In VIE as
in FDFD or related brute-force methods, leaky modes can be computed
via the solution of a generalized eigenvalue problem of the form
$Z(\omega) \xi = 0$, for a complex
frequency~\cite{marcuvitz1956field}.)  However, the FVC approach above
is advantageous in that it casts the problem in the context of
solutions of relatively few ($\ll$ degrees of freedom) scattering
calculations. More importantly, FVC can handle structures supporting
many modes or situations where near-field effects are of interest and
contribute to PE~\cite{chinamy2015near}. The latter is especially
important when the relevant quantity is the energy exchange between
two nearby objects, a regime that motivated initial development of
these and related scattering methods~\cite{liu2015near}.  Note that
above we mainly explored structures with small $\Re \epsilon \approx
4$ and large gain concentration $n$, leading to large $\Im \epsilon_g
\lesssim \Re \epsilon$ even for relatively weakly confined
resonances. However, similar effects can be obtained with smaller $n$
and $\Im \epsilon_g$ in structures with larger $\Re \epsilon$ and
dimensions, or supporting highly localized fields (e.g. spacers),
where there exist larger Purcell enhancement.

\emph{Acknowledgements} We would like to thank Li Ge, Hakan Tureci,
Steven G. Johnson, Zin Lin, and Jacob Khurgin, for useful
discussions. This work was partially supported by the Army Research
Office through the Institute for Soldier Nanotechnologies under
Contract No. W911NF-13-D-0001 and by the National Science Foundation
under Grant No. DMR-1454836.


\end{document}